\documentclass[%
 aip,
 sd,%
 amsmath,amssymb,
 reprint,%
]{revtex4-1}

\usepackage{graphicx}
\usepackage{dcolumn}
\usepackage{bm}

\usepackage[]{hyperref}
\hypersetup{colorlinks=true,linkcolor=blue,citecolor=blue,urlcolor=blue,pdfpagemode=UseNone}

\begin{document}

\preprint{AIP/123-QED}

\title{Nuclear Magnetic Resonance Probe Head Design for Precision Strain Control}

\author{T. Kissikov}
\affiliation{Department of Physics, University of California, Davis, California 95616 USA.}
\author{R. Sarkar}
\affiliation{Institute for Solid State Physics, TU Dresden, D-01069 Dresden, Germany}
\author{B. T. Bush}
\author{M. Lawson}
\affiliation{Department of Physics, University of California, Davis, California 95616 USA.}

\author{P. C. Canfield}

\affiliation{Ames Laboratory U.S. DOE and Department of Physics and Astronomy,Iowa
State University, Ames, Iowa 50011, USA}

\author{N.J. Curro}
\affiliation{Department of Physics, University of California, Davis, California 95616 USA.}

\date{\today}

\begin{abstract}
We present the design and construction of an NMR probe to investigate single crystals under strain at cryogenic temperatures.  The probe head incorporates a piezoelectric-based apparatus from Razorbill Instruments that enables both compressive and tensile strain tuning up to strain values on the order of 0.3\% with a precision of 0.001\%. $^{75}$As NMR in BaFe$_2$As$_2$ reveals large changes to the electric field gradient, and indicates that the strain is homogeneous to within 16\% over the volume of the NMR coil.
\end{abstract}

\pacs{76.60.-k, 76.60.Gv, 71.70.Fk, 77.80.bn }
\keywords{NMR, strain}
\maketitle

\section{\label{sec:level1}INTRODUCTION}

Nuclear Magnetic Resonance (NMR) is a powerful technique that probes the spin and charge degrees of freedom of matter, and is a vital tool in many branches of science including physics, chemistry, biology, materials science and medicine.  NMR is particularly important for the study of solid state materials, both in single crystals as well as polycrystalline materials, because it is versatile and can be performed in a variety of extreme environments.  To date NMR has been reported under cryogenic conditions down to sub-mK temperatures, \cite{NMRsolidHe3} at high temperatures exceeding 1500K, \cite{NMRhightemp} in static magnetic fields up to 44 T, \cite{ReyesZhengHighField}  in pulsed magnetic fields up to 62 T, \cite{HaasePulseFieldNMR} and under quasi-hydrostatic pressures up to 20 GPa.\cite{HaaseJMR2015}  Such experiments are necessary because  the behavior of matter under these extreme conditions often reveals important fundamental and practical information about the behavior of the electrons in these materials.  Recently, there has been considerable interest in the behavior of solids under strain and uniaxial stress. Strain is another physical parameter that can be used to explore new regions of phase space, particularly for metals  and for strongly correlated electron systems such as unconventional superconductors, low dimensional quantum magnets, and topological materials, in which the  Fermi surface or magnetic interaction strength can be tuned, giving rise  to strong modifications of the electronic state.\cite{Bourdarot_2011,ZieveCeCoIn52011,URSstrain2013,Sr2RuO4strainScience2014,StrainSynchrotronRSI2015,Bohmer2017PRLCa122,SternStrainSmB62017} Here we report an NMR probe head that enables NMR under tunable homogenous strain with high precision. The NMR probe is designed to operate with a Quantum Design Physical Properties Measurement System (PPMS) in fields up to 9 T and temperatures down to 1.5 K.

NMR measurements under strain have been reported previously using a horse-shoe clamp device, in which the crystal is suspended on fine wires and tensile strain is applied by tightening a screw.\cite{NMRnematicStrainBa122PRB2016}  In this case, stress on the order of a few MPa is applied at room temperature, but the strain is poorly controlled at cryogenic temperatures due to differential thermal contraction between the clamp and the crystal.  Furthermore, strain can only be applied at room temperature, requiring significant adjustments to the probe head with the possibility of misalignment.  Finally, thermal contraction may give rise to unbalanced torques and hence crystal misalignment at low temperatures.

The probe head described here is superior because it offers precision control of the strain through a combination of piezoelectric stacks and a capacitive position sensor.  This combination enables active feedback control to achieve sub-nanometer position control over time scales of several hours.

\section{Piezoelectric Apparatus}

\begin{figure}[!tb]
	\includegraphics[width=\linewidth]{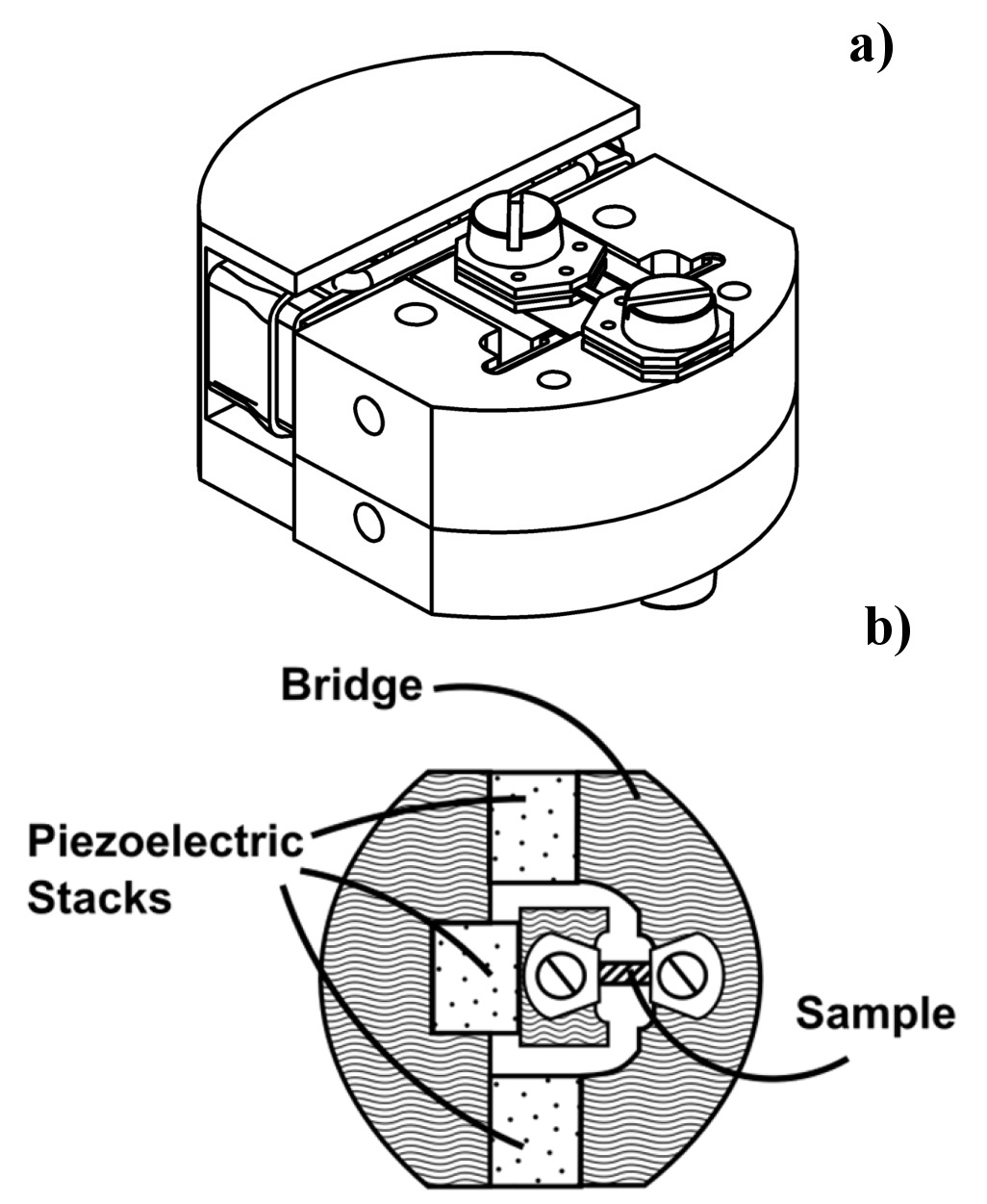}
	\caption{\label{fig:assembly} (a) Diagram of the Razorbill Instruments CS100 device, showing (b) the piezoelectric stacks and the mounted sample.}
\end{figure}

The central feature of the NMR probe head is a piezoelectric-based apparatus developed by Hicks and collaborators,\cite{Hicks2014} and commercially available from Razorbill Instruments Ltd. (Edinburgh, UK).  This device consists of a pair of piezoelectric stacks that can apply either tensile or compressive strain in situ through the application of bias voltage, as illustrated in Fig. \ref{fig:assembly}.  Because the piezoelectric stacks are arranged to cancel out thermal expansion,  strains induced by differential thermal expansions are significantly reduced, and can be overcome by applied voltage to the piezo stacks.  Our probe head, shown in Fig. \ref{fig:probehead}, integrates the CS100 device with a diameter of 25mm so that it can fit within the bore of a PPMS cryostat.

\begin{figure}[!tb]
\includegraphics[width=\linewidth]{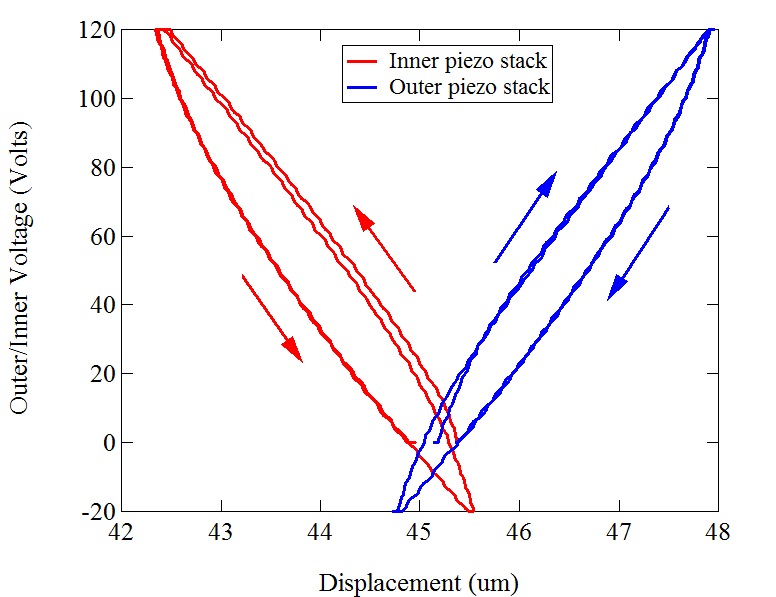}
\caption{\label{fig:hysteresis} Voltage applied to the inner (red) and outer (blue) piezoelectric stacks versus the measured displacement. Positive voltage expands the piezo stacks, leading to an expansion of the displacement for the outer stacks and a contraction of the displacement for the inner stacks. The response is hysteretic, and arrows indicate the sequence versus time.}
\end{figure}

The CS100 consists of two sets of piezoelectric stacks, one inner and two outer.  The outer stacks expand the displacement gap with positive voltage, and the inner stack contracts the displacement with positive voltage.  In tandem, the two sets of stacks can cover a range of approximately 6 $\mu$m, depending on the temperature and applied voltages, as shown in Fig. \ref{fig:hysteresis}.  The voltage is controlled via two high performance voltage amplifiers (PDm200B, PiezoDrive), and a 14-bit  USB digital-to-analog converter (USB-6001 DAQ, National Instruments) interfaced with a desktop computer.  The four voltage leads are connected via {0.8 mm	PTFE insulated copper} cables to the top of the probe. {At the top of the probe, the four voltage wires are soldered to a 4-pin panel mount hermetic LEMO connector}.

The displacement is measured by a parallel-plate capacitor integrated within the CS100 device, with a nominal spacing of approximately {40-60 $\mu$m and an area of $2.5\times 2$ mm}. {Two SC-type ultra miniature coaxial cables connected to the parallel capacitor plates are also connected to the top of the probe via hermetic panel-mount BNC connectors}. The capacitance is measured with an Andeen-Hagerling Capacitance Bridge (AH2550A), with a resolution of 0.5 attofarads at 1 kHz. This device enables us to measure the displacement with sub-nm precision.

\section{NMR Probe Head}

\begin{figure}
\includegraphics[width=0.5\linewidth]{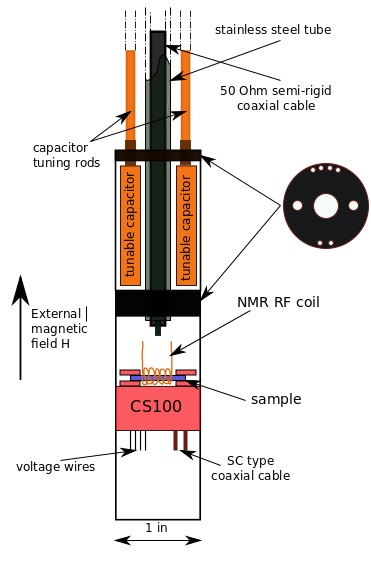}
\caption{\label{fig:probehead} Schematic diagram of the probe head. High power radiofrequency pulses are delivered to the sample via a semi-rigid coaxial cable and a tuned tank circuit using two cryogenic capacitors.  The CS100 device is mounted below at the sweet spot of the magnetic field. Four high power DC voltage wires for the piezoelectric stacks and two flexible coaxial cables for the capacitive displacement meter (not shown) go to the top of the probe.}
\end{figure}

The NMR probe consisted of a {Model 450A/B PPMS Multi-function probe} that was originally designed with a series of electrical connections at the base to connect to the PPMS electronics.  The advantage of this approach is that it enables us to easily connect electrical leads to the strained sample for resistivity measurements using the PPMS system.  The CS100 cell is mounted several cm from the base, such that the sample is located in the region of the highest field homogeneity. Two tunable cryogenic capacitors ({non-magnetic panel mount NMTM120CEK-2L, 1-120pF}, Voltronics) are mounted above the cell, as shown in Fig. \ref{fig:probehead}. These connect via a semi-rigid coaxial line to form a resonant circuit with the NMR coil. The CS100 chassis is fabricated from titanium, which has a low magnetic susceptibility and thus does not significantly affect the magnetic field homogeneity at the sample.

\begin{figure}[!tb]
\includegraphics[width=\linewidth]{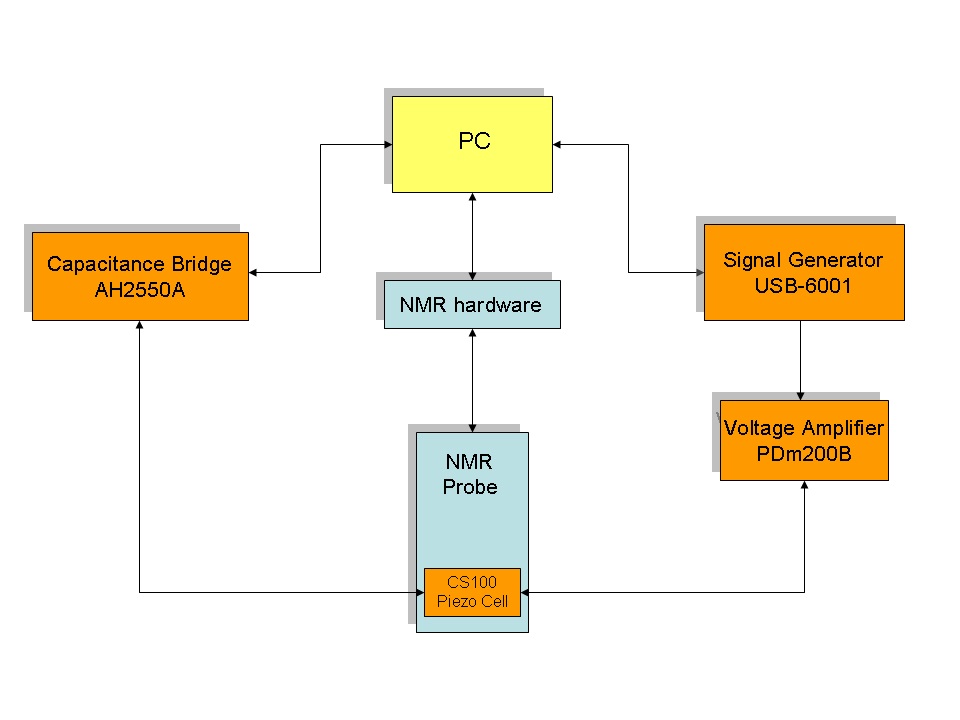}
\caption{\label{fig:diagram} The piezoelectric stacks in the strain cell are controlled by voltage amplifiers via a signal generator controlled by the computer.  The capacitive position sensor interfaces with a high precision capacitance bridge.  The voltage applied to the piezo stacks is controlled via active feedback from the computer.}
\end{figure}

An important issue for the design of the probe is the isolation of the radiofrequency coaxial cable from the high voltage piezo driver lines as well as the sensitive capacitance coaxial cables. To achieve this isolation, the radiofrequency coax passes through a stainless steel tube up to the top of the probe to hermetic BNC connections.  Shielded coaxial lines connect the rf electronics, and capacitance bridge devices to the probe, as shown schematically in Fig. \ref{fig:diagram}.

\begin{figure}[!tb]
\includegraphics[width=\linewidth]{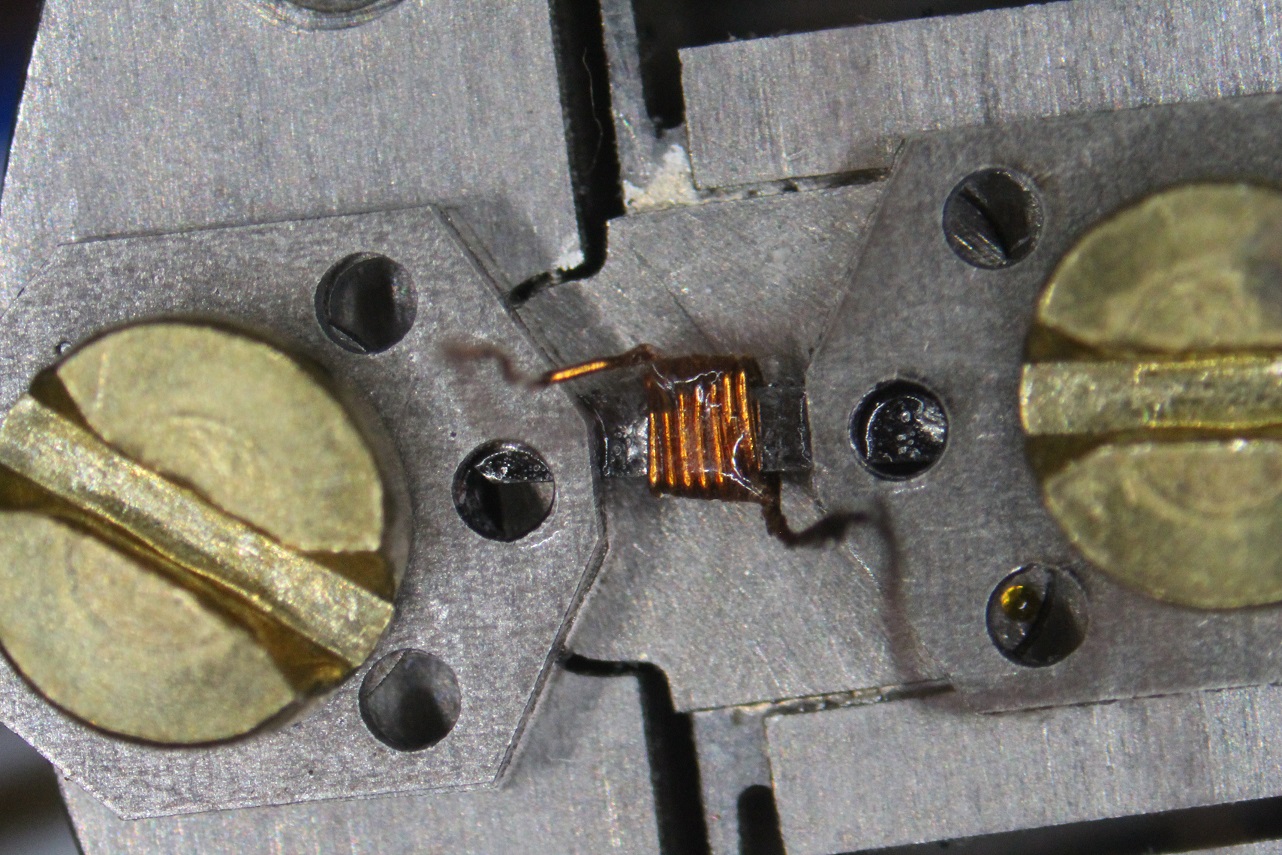}
\caption{\label{fig:coilpar} NMR coil and sample mounted for field alignment parallel to the crystalline $\hat{c}$-axis.}
\end{figure}

\begin{figure}[!tb]
\includegraphics[width=\linewidth]{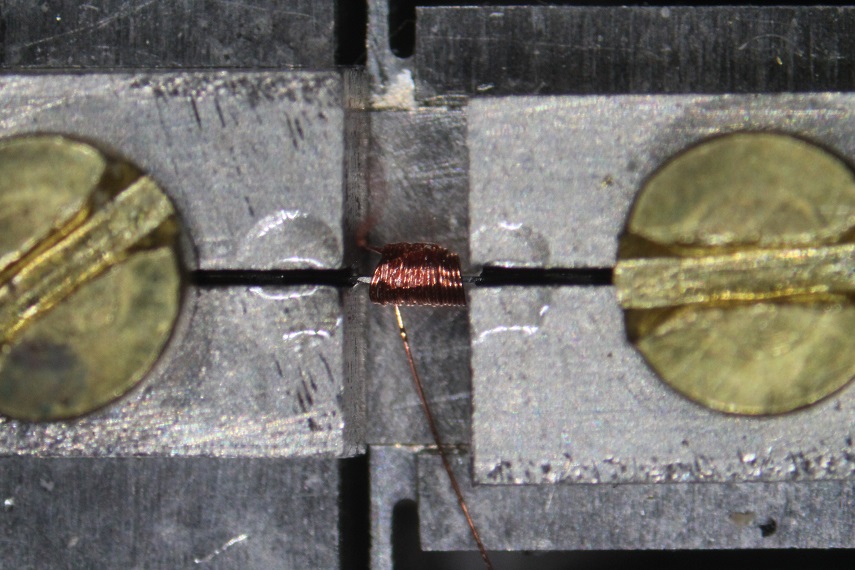}
\caption{\label{fig:fig:coilperp} NMR coil and sample mounted for field perpendicular to  the crystalline $\hat{c}$-axis.}
\end{figure}

Single crystals are cut to dimensions of approximately 0.5 mm wide by 2 mm long by 0.1mm thick. Here we report data on single crystals of BaFe$_2$As$_2$, which is known to undergo a tetragonal to orthorhombic transition at 135 K, and is well-characterized by NMR in the absence of strain.\cite{takigawa2008,NMRnematicStrainBa122PRB2016} The samples were grown and prepared as described in Refs. \onlinecite{NMRnematicStrainBa122PRB2016} and \onlinecite{NiCanfield122review}. Free standing rigid NMR coils with the appropriate inductance are placed around the sample (see Figs. \ref{fig:coilpar} and \ref{fig:fig:coilperp}), and the crystal is secured to the  CS100 strain device by epoxy (UHU Plus 300 heat-cured epoxy resin).

\begin{figure}[!tb]
	\includegraphics[width=\linewidth]{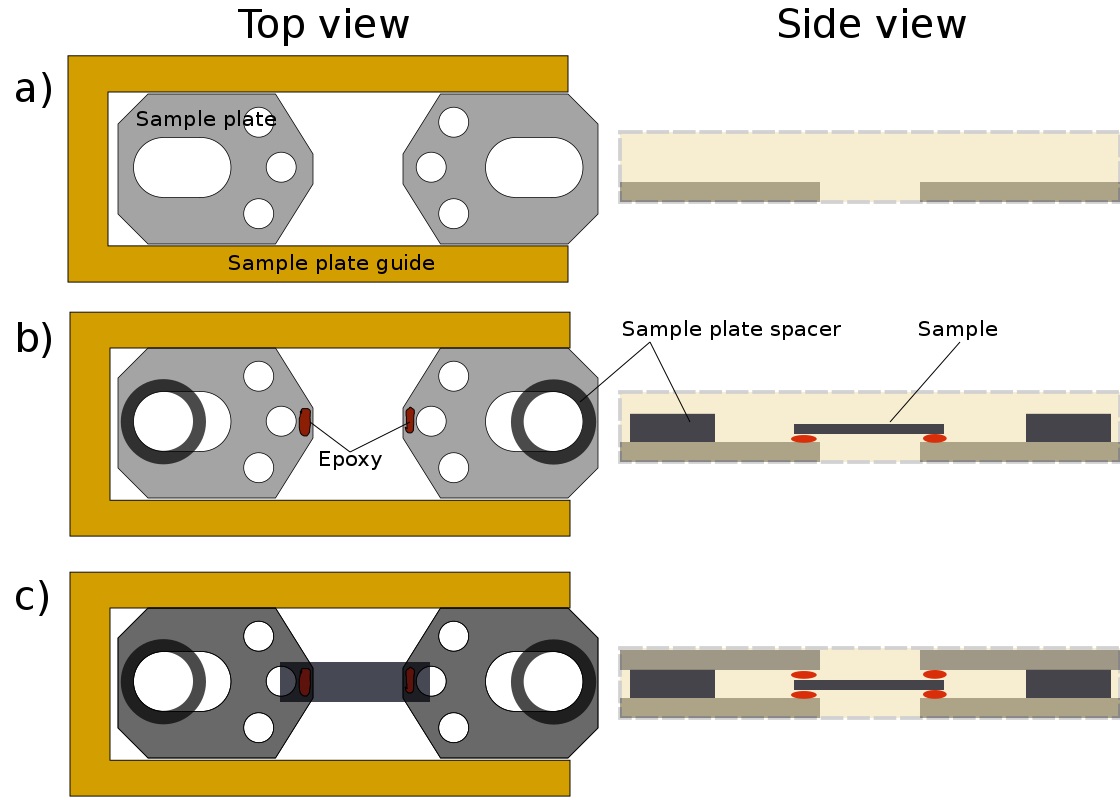}
	\caption{\label{fig:Samplemounting} Mounting sequence: (a) The sample plates are aligned with the sample plate guide; (b) The sample is secured with epoxy; and (c) the upper sample plates are secured at a distance determined by the sample plate spacers.}
\end{figure}

{To mount crystals with the magnetic field oriented parallel to the c-axis, a sample plate guide was used as shown in Fig. \ref{fig:Samplemounting}. First, two sample plates were fastened to the cell using M2 brass screws and aligned parallel using the sample plate guide, which itself is fastened to the cell using M1.6 screws. The distance between the two mounting pieces along the strain axis were chosen to ensure that the NMR coil fits in between and sits freely. A drop of UHU 300 heat-cure epoxy is then deposited to the two ends of the sample plates using a thin stainless steel wire. In principle, two thin wires should be placed along with epoxy so that crystal is aligned parallel to the cell's surface. After the sample is secured, another layer of epoxy is deposited to the upper surface of the sample. Sample plate spacers, which have been polished down to limit the epoxy thickness to 30-50 $\mu m$ on both sides of the crystal, are placed on both the sample plates. Finally, the upper sample plates are placed above the sample plate spacers and epoxy, and fastened to the cell using M2 brass screws. The cell with the sample is then placed under a heat lamp to let the epoxy cure at around 75-80 $^{\circ}$C for 40 minutes.}

{For crystals mounted with the field oriented perpendicular to the c-axis, we constructed a mounting device by cutting a thin vertical slot in two blocks of (grade 2) titanium of dimensions $~5\times5.6\times1.4$ mm$^3$ as shown in Fig. \ref{fig:fig:coilperp}.
Slits of width $\sim0.2$ mm were cut using a High-Speed Steel Slitting Cutter (0.008 inches thick). Holes were tapped to secure the two mounting pieces on the cell using the M2 screws. The two slits were aligned along the strain axis under a Nikon SMZ800 microscope, and with the modified mounting pieces we can use the sample plate guide.  Before placing the crystal and NMR coil, a small amount of UHU Plus 300 heat-cured epoxy was deposited into the slits. After the crystal with the free standing NMR coil was placed, more epoxy was deposited into the slits to fill the space between the crystal plates and the walls in the slits. The slits are sufficiently narrow and the viscosity of the epoxy was sufficiently high that we did not encounter any problems with the epoxy flowing out of the slits on the open sides. Nevertheless, the epoxy was filled in over a sequence of several small steps, and allowed to cure each time.}

In principle, the CS100 can also be rotated by 90 degrees, however in this case the strain axis is vertical, parallel to the applied field.  This orientation is unacceptable, though, because the NMR rf field, $\mathbf{H}_1$, must be perpendicular to the applied field in order to induce a signal. Therefore, a solenoidal coil around the sample that is oriented along the strain axis would not work.  We have had some success with a Helmholz coil design, which would enable the field to be oriented along the strain axis, while $\mathbf{H}_1$ is perpendicular to the applied field.  However, the signal-to-noise for this orientation was significantly reduced. Alternatively, a rigid coil that is larger than the sample, but oriented such that $\mathbf{H}_1$ has a component perpendicular to $\mathbf{H}_0$ can be used.\cite{NMRnematicStrainBa122PRB2016} Note, however, that it is possible to investigate the in-plane anisotropy without orienting the strain axis along $\mathbf{H}_0$.  By symmetry, negative (compressive) strain with $\mathbf{H}_0$ perpendicular to the strain-axis is approximately equivalent to positive (tensile) strain with $\mathbf{H}_0$ parallel to the strain axis.  For materials with a sufficiently large Poisson ratio there are strains introduced along the directions perpendicular to the strain axis, and other strain modes such as $\epsilon_{xx} + \epsilon_{yy}$ may be present. As a result, the equivalence between the field directions and a real rotation of the sample is broken, which could produce asymmetries in quantities measured via this method. However, we have found linear behavior over a broad range of strains in the BaFe$_2$As$_2$ sample studied here, suggesting that these other modes do not play a significant role in this case.\cite{Kissikov2017}

\section{Active Feedback Control}

\begin{figure}[!tb]
\includegraphics[width=\linewidth]{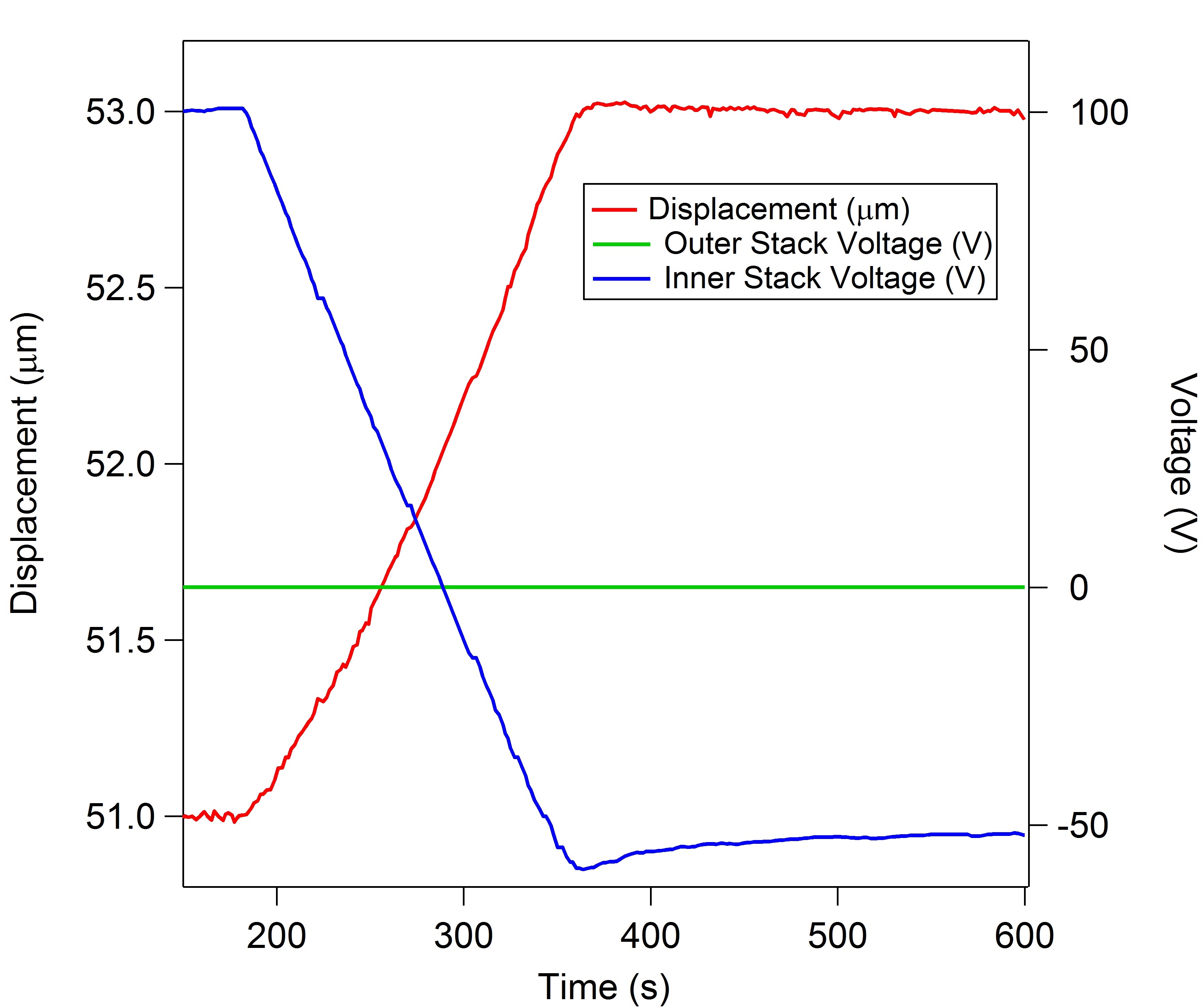}
\caption{\label{fig:responsevstime} Displacement (red), inner (blue) and outer (green) stack voltages versus time. At approximately 190 s, the setpoint displacement was set to 53.0 $\mu$m.
The setpoint was reached and stabilized by approximately 180 s.  The applied voltage drifted over time as the active feedback compensated for the drift of the piezoelectrics.  }
\end{figure}

The creep behavior of the piezoelectric actuators, in which the length changes at constant voltage, presents a challenge for long-term NMR measurements over the course of several hours. After setting a new voltage or temperature, the displacement of the stacks drifts by several percent over the course of several minutes to hours.  If the NMR properties under investigation are strain-sensitive, this creep can give rise to unacceptable errors.  Long-term stability is also important for other measurements, such as specific heat, $\mu$SR, and neutron scattering where time duration is an important factor.  For example, at large-scale facilities with limited beam time availability, waiting for the piezoelectric stacks to relax would be prohibitively expensive. In order to overcome these issues, we implemented active feedback control using a PID control loop.  A Python script running on the computer uses the input signal from the capacitance bridge and determines an error based on a desired displacement setpoint.  The output voltage is then determined based on this error signal.  Depending on the setpoint, either the inner or the outer stack is used for the feedback control, while the other set of stacks is left to drift at a fixed voltage.  Fig. \ref{fig:responsevstime} shows the time-dependence of the displacement and voltage after a new setpoint is implemented. 
Using this approach, we have been able to achieve displacements that stabilize quickly and remain stable over days with 0.9 nm rms fluctuations.

\section{Electric Field Gradient Control}

\begin{figure}[!tb]
\includegraphics[width=\linewidth]{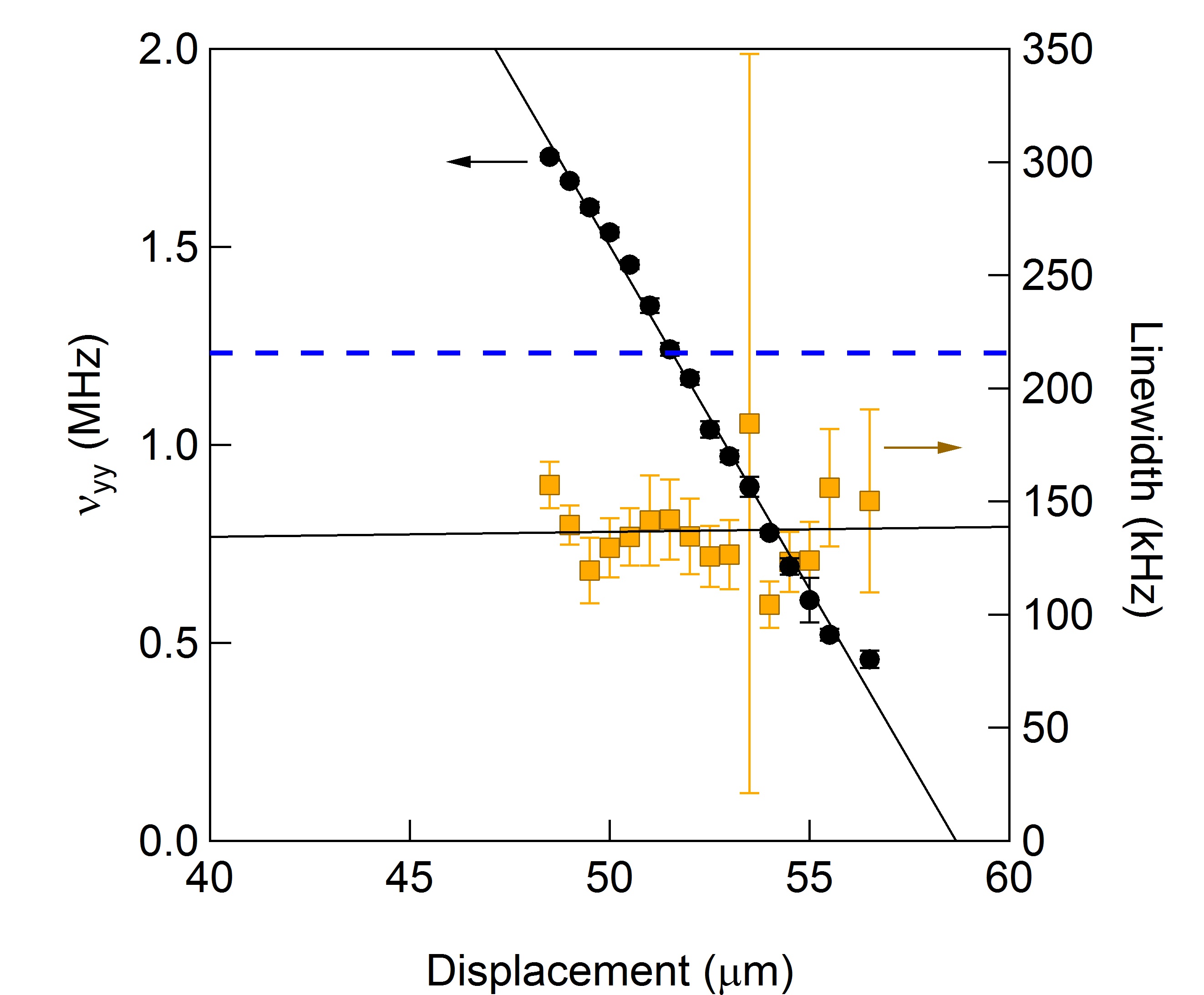}
\caption{\label{fig:EFGvsDisplacement} Quadrupolar splitting and quadrupolar linewidth versus displacement for the As site in BaFe$_2$As$_2$ at 138K. The solid line is a linear fit to the data, and the dashed line is the value in the zero-strained case.}
\end{figure}

Determining the strain, $\epsilon = (x-x_0)/L_0$, where $x$ is the displacement and $L_0$ is the sample length, requires knowledge of the unstrained length, $x_0$.  However,  differential thermal contraction at cryogenic temperatures will give rise to finite strains, even if the crystal is secured at room temperature in zero strain.  Although it is possible to employ a strain gauge to calibrate the displacement, a superior approach is to measure an intrinsic NMR property of the material of interest.  An ideal quantity is the electric field gradient (EFG) asymmetry of a quadrupolar nucleus.  Here we consider the EFG of $^{75}$As ($I=3/2$) in BaFe$_2$As$_2$, shown in Fig. \ref{fig:EFGvsDisplacement} at 138K.   Above the structural transition, the EFG tensor at the As site has tetragonal symmetry, and therefore $\nu_{xx} = \nu_{yy} = -\nu_{zz}/2$.\cite{CPSbook} Strain, however, breaks the tetragonal symmetry and gives rise to a non-zero asymmetry parameter, $\eta = (\nu_{xx} - \nu_{yy})/\nu_{zz}$. In this case the quadrupolar splitting, $\nu_{yy}$ is a strong function of the displacement, as shown in Fig. \ref{fig:EFGvsDisplacement}.   The strong linear variation indicates that the asymmetry parameter is changing with displacement. By comparing $\nu_{yy}$ with the value in the unstrained case, we identify $x_0 = 51.53$ $\mu$m as the displacement in the absence of strain at this temperature, where $\eta=0$.

Although the epoxy securing the crystal to the device may be expected to deform, the  strong linear variation of the EFG with strain indicates that the a large portion of the strain is indeed transferred to the sample.
Furthermore, we find that the linewidth of the quadrupolar satellite does not exhibit any significant variation with strain, as shown in Fig. \ref{fig:EFGvsDisplacement}.  This result indicates that the strain remains homogeneous over the volume of the crystal within the NMR coil, otherwise we would observe broadening of the resonance with increasing strain.  We observe that $d\nu_{yy}/d\epsilon = -259$ MHz, whereas the average linewidth is $\delta \nu_{yy} = 125$ kHz, implying an upper bound on the  strain inhomogeneity of less than $\delta \epsilon \lesssim 4.8\times 10^{-4}$.  The strain varies from -0.002 to +0.003 over this range, so $\delta \epsilon/\epsilon \lesssim 16$\%.

\section{Conclusions}

The NMR probe with precision strain control enables a new regime of NMR experiments under extreme conditions.  Homogeneous displacements up to $\pm 6$ $\mu$m at room temperature ($\pm 3$ $\mu$m at cryogenic temperatures) can be achieved with a stability of 0.5 ppm over multiple hours for a sample length of $\sim 2$ mm.  NMR quantities such as the Knight shift, EFG, and spin-lattice relaxation tensors can be probed, which can shed new and important light on the electronic behavior of materials under extreme conditions.  In light of these new results, it may be worthwhile to revisit previously studied materials using strain to tune the electronic properties.

\begin{acknowledgments}
We thank A. Dioguardi, M. Tanatar R. Fernandes and I. Fisher for enlightening discussions, K. Shirer,   L. Simon, and  P. Klavins for assistance in the laboratory, and acknowledge discussions with A. Ward from Razorbill Instruments. Work at UC Davis was supported by the NSF under Grant No.\ DMR-1506961 and the NNSA under Grant No. DE-NA0002908. R. Sarkar was partially supported by the  DFG through SFB 1143 for the project C02. Work done
at Ames Lab was supported by the U.S. Department
of Energy, Office of Basic Energy Science, Division of Materials Sciences
and Engineering. Ames Laboratory is operated for the U.S. Department
of Energy by Iowa State University under Contract No. DE-AC02-07CH11358.
\end{acknowledgments}

\bibliography{NMRStrainProbe}

\end{document}